\def\year{2021}\relax
\documentclass[letterpaper]{article} 
\usepackage{aaai21}  
\usepackage{times}  
\usepackage{helvet} 
\usepackage{courier}  
\usepackage[hyphens]{url}  
\usepackage{graphicx} 
\usepackage{natbib}  
\usepackage{caption} 

\usepackage[usenames,dvipsnames,table]{xcolor}
\usepackage{verbatim}
\usepackage{url}
\usepackage{bm}
\usepackage{booktabs}
\usepackage{multirow}
\usepackage{amssymb}
\usepackage{amsmath}
\usepackage{soul}

\newcommand{\ct}[1]{\citeauthor{#1} \shortcite{#1}}

\title{Towards Fairness in Classifying Medical Conversations into SOAP Sections}
 \author{
     Elisa Ferracane,\textsuperscript{\rm 1}
     Sandeep Konam \textsuperscript{\rm 1}
     \\
 }
  \affiliations{
     \textsuperscript{\rm 1}Abridge AI Inc.\\

     elisa@abridge.com, san@abridge.com

 }

\begin{document}

\maketitle

\begin{abstract}
As machine learning algorithms are more widely deployed in healthcare, the question of algorithmic fairness becomes more critical to examine. Our work seeks to identify and understand disparities in a deployed model that classifies doctor-patient conversations into sections of a medical SOAP note. We employ several metrics to measure disparities in the classifier performance, and find small differences in a portion of the disadvantaged groups. A deeper analysis of the language in these conversations and further stratifying the groups suggests these differences are related to and often attributable to the type of medical appointment (e.g., psychiatric vs. internist). Our findings stress the importance of understanding the disparities that may exist in the data itself and how that affects a model's ability to equally distribute benefits.
\end{abstract}

\begin{table*}[h!]
\begin{center}
\begin{tabular}{llp{7.7cm}}
\toprule
Metric &Formula &Thresholds\\
\midrule
\texttt{AOD}: average odds difference &$\frac{(FPR_{\text{disadv}} - FPR_{\text{adv}}) + (TPR_{\text{disadv}} - TPR_{\text{adv}})}{2}$ &equal: 0, $<$-0.1: higher benefit for advantaged group, $>$0.1: higher benefit for disadvantaged group  \\
\texttt{EOR}: equal opportunity ratio & $\frac{TPR_{\text{disadv}}}{TPR_{\text{adv}}}$ &equal: 1, $<$0.8: higher benefit for advantaged group, $>$ 1.25:  higher benefit for disadvantaged group \\
\texttt{FORR}: false omission rate ratio & $\frac{FOR_{\text{adv}}}{FOR_{\text{disadv}}}, 
 FOR = \frac{FN}{FN + TN} $ &equal: 1, $<$0.8: higher benefit for advantaged group, $>$ 1.25: higher benefit for disadvantaged group\\
 \bottomrule
\end{tabular}
\caption{{\small Group disparity metrics. FPR=false positive rate; TPR=true positive rate; FOR=false omission rate.}}
\label{tab:fair_metrics}
\end{center}
\end{table*}

\section{Introduction}
Disparities in healthcare in the U.S. have existed long before machine learning algorithms were introduced. An early seminal report from the Institute of Medicine (IOM) concludes racial and ethnic minorities receive poorer healthcare 
 \cite{Smedley:2003}, a finding still observed in recent work and extending to other disadvantaged groups \cite{Jusot:2019,Williams:2019}.

Applying machine learning solutions to healthcare problems holds great promise, but also poses challenges to understand disparities that it may introduce or amplify \cite{Chen:2020}. For example, \ct{Obermeyer:2019} examine an algorithm which identifies high-need patients in order to allocate more medical resources for them. The algorithm uses cost of care as a proxy for high-need, which the authors show is a poor and unfair construct since Black patients with the same cost of care (as Whites) have greater health needs but are being denied the benefit of the program.

Most machine learning solutions in healthcare are based on metrics for medical interventions and outcomes often sourced from electronic health records \cite{Chen:2020}. Considerably less work makes use of doctor-patient conversations, even though these are known to play an important role in patient health \cite{Ong:1995}. 
Our work is the first to focus on the algorithmic disparities of a classifier that operates over doctor-patient conversations.
 
We examine a classifier that categorizes utterances of a medical conversation into different sections of a SOAP note (see example Figure \ref{fig:note_example} in the Appendix). SOAP is an acronym for Subjective, Objective, Assessment, and Plan, referring to the four major sections of the problem-oriented medical note. A SOAP note is produced by a medical provider in order to summarize an encounter with a patient. The Subjective section includes information reported by the patient (such as past illnesses, current symptoms), the Objective consists of what the provider measures and observes (patient's temperature, an EKG), the Assessment is the provider's diagnosis of the patient (e.g., recovering well), and the Plan is the provider's plan for treating the problem (which can include drugs and therapeutics or further tests and appointments). In addition to the above 4 classes, we consider `None' as the fifth class to denote utterances not relevant to the SOAP note, such as chit-chat. Output of this classifier can improve recall and understanding
of care plans for patients \cite{Schloss:2020}. The Plan section is the most important of all sections from a patient adherence standpoint, as it can enable patients to better follow through on their care (e.g., to fill out a new prescription, or schedule a new appointment). 

Motivated by concerns of fairness for the distribution of this patient benefit, we analyze the performance characteristics of the SOAP note classifier in order to detect possible disparities towards disadvantaged groups. Guided by the IOM report \cite{Smedley:2003} and recent healthcare studies on unequal treatment, we examine 7 protected attributes and define 18 disadvantaged groups (including intersectional ones that cross multiple protected attributes).

Using three carefully selected metrics most appropriate for our scenario, our findings show small but statistically significant disparate outcomes for fewer than half of the disadvantaged groups examined, as measured by one of the metrics. Probing deeper into these groups, we find differences in the language used in their conversations and what kind of medical provider these groups are visiting (e.g., psychiatrist vs. internist). These results suggest the disparate outcomes often stem from the type of medical visit, and underline the importance of understanding how disadvantaged groups differ in their use of medical care from the general population \cite{Collins:1999,Hegarty:2000}.  

\begin{table}[!h]
\begin{center}
\begin{tabular}{lcccp{1.5cm}}
\toprule
SOAP section & \# utterances (\%) &in conv &F1\\
\midrule
Subjective & \phantom{1,}190,914 (16.6\%)&4,975  &51.3\\ 
Objective & \phantom{11,}23,741 \phantom{0}(2.1\%)&2,962 &38.7 \\
Assessment & \phantom{1,}175,326 (15.3\%)&4,830 &31.6 \\
Plan & \phantom{11,}38,130 \phantom{0}(3.3\%)&3,999 &24.5 \\
none &\phantom{1,}720,478 (62.7\%)&4,992  &27.7 \\
\midrule
Total &1,148,589 &5,000 &37.6\\
\bottomrule
\end{tabular}
\caption{{\small Statistics and classifier performance on the validation set.}}
\label{tab:data_stats}
\end{center}
\end{table}

\section{Data and Classifier}

The dataset consists of 63,000 doctor-patient conversations that have been collected with full consent, de-identified, and transcribed by humans. The data is split into 52,000/5,000/6,000 for train/validation/test. Our work focuses exclusively on the validation set, summarized in Table \ref{tab:data_stats}. The dataset includes sociodemographic data about the patient, the physician, and the location of the medical facility, which we detail in the following section.\footnote{The metadata is collected by the medical provider.}

We use a classifier based on \ct{Schloss:2020} 
which classifies each utterance into one of five classes in accordance with the SOAP note: Subjective, Objective, Assessment, Plan, or none of these. The classifier encodes each utterance using ELMo word embeddings \cite{Peters:2018}, attention and a stacked bi-LSTM. A decoder LSTM then predicts the class for the utterance. 

\section{Evaluation of Group Disparities}
Group disparities exist when outcomes differ systematically between population groups \cite{Barocas:2019}. A group disparity could lead to allocative harm, by withholding benefit, or representational harm, by reinforcing the subordination of a group \cite{Crawford:2017}. The SOAP classifier provides potential benefits to patients by highlighting important parts of their conversation (usually utterances marked as Plan) in order to better follow through on their care. With an eye towards fairness, our work focuses on possible allocative harm: whether these benefits are withheld from  disadvantaged groups. 

We measure disparities between groups under the (naïve) assumption that the dataset distribution reflects the true population distribution. We recognize this is likely not the case and, further, that biases have likely been introduced earlier in the dataset creation process, starting from our choice of variable (examining medical conversations disfavors groups with fewer medical visits stemming from cultural or economic reasons, such as distrust of doctors or risk of job loss). We choose to focus on the disparities of the classifier as a starting point because we have control over the model. In future work, we intend to examine other sources of bias.

We view our classification task as assistive and we thus choose three metrics focused on true positives and false negatives (Table \ref{tab:fair_metrics}): average odds difference (\texttt{AOD}), equal opportunity ratio (\texttt{EOR}), false omission rate ratio (\texttt{FORR}).
For the last two metrics, we compute the ratio between the two groups (as opposed to the difference) in order to better capture smaller differences. We focus on false omission rate instead of false negative rate because the former captures the conditional probability of the ground truth based on the predicted outcome, which is known to the decision maker, while latter is conditioned on the usually unknown ground truth \cite{Rodolfa:2020,Mitchell:2018}.

To judge whether the magnitude of a difference is disparate, we adopt the thresholds in \ct{ZhangY:2020}, which are based on legal guidelines from the U.S. Equal Employment Opportunity Commission. Statistical significance is determined using a permutation test as detailed in \ct{DiCiccio:2020} which calculates the studentized metric on all possible permutations of the data split into two groups (approximated by shuffling the data N=1000 times). The null hypothesis assumes the two populations are equal, which we reject at $\alpha<0.01$.  

The groups we choose to focus on are based on recommendations from the IOM report 
and prior work showing bias against these groups specifically in the medical field.  See Table \ref{tab:fair_groups} in the Appendix for coverage of the metadata (not every conversation has all the attributes studied) and detailed group statistics. We next explicitly define the groups within each protected attribute, and give a brief overview of medical bias against the groups.

 
\medskip
\noindent
\textbf{Race/Ethnicity} The dataset includes a single field for both race and ethnicity. We recognize this protected attribute conflates two notions which are both ill-defined and subject to reporting bias 
\cite{Kressin:2003}. We define one advantaged group (`White') and three disadvantaged groups: `Black', `Hispanic', `Asian'. 

Historically, Black patients are less likely to be recommended for operations, less likely to receive treatments and more likely to die \cite{Ford:1989,Lafata:2001,Howard:1998}. At the same time, algorithmic bias against Black patients has surfaced by both causing harm (language models recommending ``prison'' instead of ``hospitals'' for violent Black patients; \citeauthor{ZhangH:2020} \citeyear{ZhangH:2020}) and by denying benefit (excluding Black patients from a beneficial health program by using health costs as a discriminatory proxy for health needs; \citeauthor{Obermeyer:2019} \citeyear{Obermeyer:2019}). 

Hispanic and Asian patients face many of the same medical disadvantages as Blacks \cite{Ramsey:1997,Hannan:1999,Carlisle:1995,Thamer:2001}, in addition to linguistic barriers for patients with limited or low English proficiency and the lowest insurance coverage among minorities often as a result of immigration status \cite{Smedley:2003}. Algorithms that predict outcomes such as disease susceptibility using electronic health record data fail to account for these confounding factors \cite{Gianfrancesco:2018}. Furthermore, word embeddings used pervasively in NLP-based healthcare algorithms are shown to capture ethnic stereotypes and biases \cite{Garg:2018}.

\medskip
\noindent
\textbf{Gender} The reported gender in the dataset is limited to `male' or `female', and does not include transgender or non-cisgender choices. We define the advantaged group as `male' and the disadvantaged as `female'. Prior work shows women experience longer delays in diagnosis \cite{Lyratzopoulos:2013}; doctors take the pain of women less seriously, often attributing it to psychogenic causes \cite{Samulowitz:2019}. Medical-domain word embeddings, which can be helpful for medical NLP tasks, nevertheless also reflect these gender biases
\cite{Rios:2020}.

\medskip
\noindent
\textbf{Socio-economic status} The dataset includes socioeconomic information about the patient: employment status (the advantaged group is `full-time job', disadvantaged are `unemployed' and `retired'),  living circumstance (advantaged is `living at home', disadvantaged are `nursing home' and `incarcerated'), and medical insurance (advantaged is `private insurance', disadvantaged are `Medicaid', `uninsured'). Lack of employment and insurance are associated with poorer healthcare, and are more prevalent among ethnic and racial minorities \cite{Smedley:2003}. Living in a nursing home has shown mixed results for patient health. Studies show improved medical quality of life and social support for those in nursing homes, but also increased mortality risk when compared to living at home \cite{Stevens:2014}.  Patients who are incarcerated are more likely to receive fewer treatment options and overall lower quality of healthcare \cite{Fuller:2017}.

\medskip
\noindent
\textbf{Age} The dataset includes patient age, ranging from 0-98 years old.
We bin the ages into three groups: youth (0-17), adult (18-64), older adult (65-98). The advantaged group is `adult', and disadvantaged are `youth' and `older adult'. Less money is invested in children's health despite often  poorer outcomes \cite{Leatherman:2004}, while older adults are less likely to receive kidney transplants and treatment for pain \cite{Kjellstrand:1988,Bernabei:1998}.

\medskip
\noindent
\textbf{Obesity} The dataset reports the weight of the patient, which we use as a crude proxy for obesity ($>=$250lbs. is the disadvantaged obese group). Studies show doctors often have explicit and implicit biases against overweight patients, and the patients experience poorer care and worse outcomes \cite{Sabin:2012,Tomiyama:2018}.

\medskip
\noindent
\textbf{Mental health} The dataset reports the physician's specialty, which we use as a proxy for patients with mental health issues. The disadvantaged group is patients seeing psychiatrists, and the advantaged group is those seeing all other physicians. Mental illness patients are often subject to `diagnostic overshadowing' where physical illness signs are misattributed to their psychological disorder, resulting in poorer quality of care \cite{Thornicroft:2007}.

\medskip
\noindent
\textbf{Location} The dataset includes the U.S. state where the medical visit occurred. A recent study uncovered many machine learning algorithms are trained on a geographically limited cohort of patients (Florida, California and New York; \citeauthor{Kaushal:2020} \citeyear{Kaushal:2020}). Because many health conditions correlate with geographic location, this imbalance could result in unfair treatment for patients from other areas. The dataset we examine is not as geographically restricted (45 of the 50 states are included), but is slightly skewed (the three states comprise a third of the data).

\medskip
\noindent
\textbf{Intersectional} We further explore groups that intersect multiple protected attributes. Specifically, we examine Black females and Hispanic females, based on prior work showing these groups experience poorer care and outcomes compared to White males \cite{Smedley:2003}. 

\section{Results}
We evaluate 18 groups across 5 SOAP note sections (total 90 combinations) using the metrics and thresholds in Table \ref{tab:fair_metrics}, and report results only on groups that exceed the thresholds in favor of the advantaged group (see Tables \ref{tab:all_results_1} and \ref{tab:all_results_2} in the Appendix for all results). As shown in Table \ref{tab:results}, we find 7 cases (out of 90) with a statistically significant higher benefit for the advantaged group. Importantly, we note that in all these cases, only one of the three metrics reaches the disparate threshold. In datasets where unfairness is well-established, the algorithms exceed the disparate thresholds on multiple metrics \cite{Bellamy:2018}. We next analyze each group in more detail.

\begin{table*}[!t]
\begin{center}
\scalebox{0.98}{
\begin{tabular}{p{1.1cm}p{2.1cm}p{2.3cm}p{2.1cm}p{1cm}p{1cm}p{1cm}p{4cm}}
\toprule
SOAP section &Protected attribute & Disadvantaged group &Advantaged group &\texttt{AOD} \small{$[\text{-}0.1,0.1]$} &\texttt{EOR} \small{$[0.8,1.25]$} &\texttt{FORR} \small{$[0.8,1.25]$} &new\phantom{'}\texttt{FORR}\phantom{'}(omitted\phantom{'}medical \phantom{aaaaaa'''''''''''}providers)\\
\midrule
\multirow[t]{3}{*}{Plan} & Race      &Asian  &White &\textcolor{ForestGreen}{\phantom{-}0.00} &\textcolor{ForestGreen}{0.97} &\textcolor{red}{0.69} &\textcolor{ForestGreen}{0.83}\phantom{''}(clinical\phantom{'}cardiologist, \phantom{000'''} ophthalmologist) \\
& Race+Gender      &Hispanic female  &White male &\textcolor{ForestGreen}{-0.01} &\textcolor{ForestGreen}{0.95} &\textcolor{red}{0.71} &\textcolor{ForestGreen}{0.82}\phantom{''}(allergist)  \\
&Socioeconomic &Incarcerated &Living at home &\textcolor{ForestGreen}{-0.05} &\textcolor{ForestGreen}{0.95} &\textcolor{red}{0.71} &\textcolor{ForestGreen}{0.82}\vspace{0.2cm}\phantom{''}(inf. disease specialist)\\
\multirow[t]{2}{*}{Objective} &Socioeconomic &Uninsured
&Private insur. &\textcolor{ForestGreen}{-0.02} &\textcolor{ForestGreen}{0.94} &\textcolor{red}{0.67} &\textcolor{blue}{1.71}\phantom{''}(inf. disease specialist)\\
&Age &Older adult &Adult &\textcolor{ForestGreen}{\phantom{-}0.00} &\textcolor{ForestGreen}{0.99} &\textcolor{red}{0.69} &\textcolor{red}{0.75}\phantom{''}(ophthalmologist) \vspace{0.2cm}\\
\multirow[t]{2}{*}{None} &Race &Black &White &\textcolor{ForestGreen}{-0.04} &\textcolor{red}{0.78} &\textcolor{ForestGreen}{1.01} &-\\
&Race+Gender &Black female &White male &\textcolor{ForestGreen}{-0.04} &\textcolor{red}{0.76} &\textcolor{ForestGreen}{1.01} &-\\
\bottomrule
\end{tabular}}
\caption{{\small Groups with higher benefit for the advantaged group (as measured by at least one metric). Numbers in green are within the thresholds, red are disparate favoring the advantaged group, blue favoring the disadvantaged group. The new \texttt{FORR} is recalculated after omitting the listed medical provider type. \texttt{AOD}=average odds difference, \texttt{EOR}=equal opportunity ration, \texttt{FORR}=false omission rate ratio.}}
\label{tab:results}
\end{center}
\end{table*}

\section{Analysis}
To understand whether the disparities could be related to differences in the language of the doctor-patient conversations, and whether the differences are attributable to other confounding factors, we conduct two analyses. First, we perform a conditional word frequency analysis to identify lexical cues that are strongly indicative of the class, but that are absent from the disadvantaged group conversations. Specifically, we calculate the local mutual information (LMI) between the $n$-grams of the conversations and the class, and hypothesize the classifier would perform more poorly if the top-scoring  $n$-grams in the overall conversations are absent (or much less prominent) in the disadvantaged group conversations. Second, because differences in conversation are expected and warranted in different types of medical visits, we experiment with omitting visit types from the group and observe the effect on the classifier error rates.

We first discuss disparities for the `Plan' section, which provides the greatest benefit to the patient. In the Asian group with more false omissions, the LMI analysis shows `blood' and `blood work' are strong lexical cues for the class, but are less prominent in the Asian conversations, suggesting a different distribution of medical visit types. Indeed, visits to the clinical cardiologist and ophthalmologist are more frequent in the Asian group, though they comprise a small proportion (5.3\% of Asian `Plan' utterances vs. 3.6\% of White, and 1.0\% vs. 0.2\%, respectively). If we omit these cases from the Asian group, the \texttt{FORR} no longer exceeds the thresholds, as seen in the last column of Table \ref{tab:results}. Furthermore, when comparing these two specialties to all other specialties across the entire population, we find these two are inherently and significantly harder for the model to classify, regardless of race or ethnicity. 

For the Hispanic female group with more false omissions, the LMI analysis shows results similar to the Asian group: lexical cues in the overall conversations are less prominent in the Hispanic ones ($n$-grams `blood' and `blood work'). We again explore the type of medical appointment and find omitting the more common visits to the allergist (5.1\% vs. 2.3\% in White) eliminates the observed disparities. However, we find the allergist visits are harder to classify only within the Hispanic group, and not for the entire population. 

Incarcerated patients experience more false omissions, and the LMI analysis suggests a wider gap in the nature of the conversations: in addition to `blood' and `blood work' attested in other groups, other $n$-grams suggestive of future appointments are also less prominent (e.g., `next', `months', `'re going',). Omitting the slightly more common visits to the infectious disease specialist (5.7\% vs. 1.9\% for patients living at home) eliminates the disparities. However, we do not observe this specialty is harder for the model to classify across the two living condition groups.

We next explore errors in the `Objective' section, although we note there are fewer benefits and thus lower harm from potential disparities. In the uninsured patient group with more false omissions, the LMI analysis reveals results similar to the Asian group: `blood' is a strong lexical clue but is less prominent in the uninsured group. We again find visits to the infectious disease specialist are more common for uninsured vs. privately insured patients (6.3\% vs. 1.7\%), and omitting these cases erases the disparities. We also observe these types of visits are inherently harder for the model to classify, regardless of insurance status.

The older patient group with more false omissions has a more ambiguous LMI analysis, where `liver' is less important. Omitting the internist and ophthalmologist visits mitigates but does not eliminate the disparities, and we do not find these visits to be inherently harder across all ages. 

For the `none' section, we posit little to no benefit for the patient when utterances are correctly categorized as chit-chat. 
We find fewer true positives for Black and Black female groups, with no differences in the LMI analysis, showing lexical cues are similar across groups. Because there is no clear harm, we do not pursue these differences.

In summary, our analysis finds considerable variation in the conversations of the different disadvantaged groups. These differences are often related to the types of medical visits and sometimes the type of visit is the true factor underlying the group disparities. 

\section{Conclusion}
Motivated by concerns for fairness, we analyze the performance characteristics of a classifier that categorizes doctor-patient conversations in order to identify and understand disparate outcomes for disadvantaged groups. Our results show small but statistically significant differences for a portion of the groups, though only as measured by one of the three metrics. We further analyze the conversation within these disadvantaged groups to find variations that can be traced back to different types of medical visits. The type of visit can sometimes entirely account for the attested disparities. This finding highlights the importance of understanding the differences already present in datasets, and how these can affect a model's ability to allocate equal benefit. 
In future work, we aim to understand the longer term effects of disparities, as in the simulated studies of \ct{DAmour:2020}. 

{\small
\bibliography{aaai21}}
\bibliographystyle{aaai21}

\section{Appendix A: Example}
\label{sec:appendix_a}

\begin{figure*}[!h]
\centering
\includegraphics[scale=0.25]{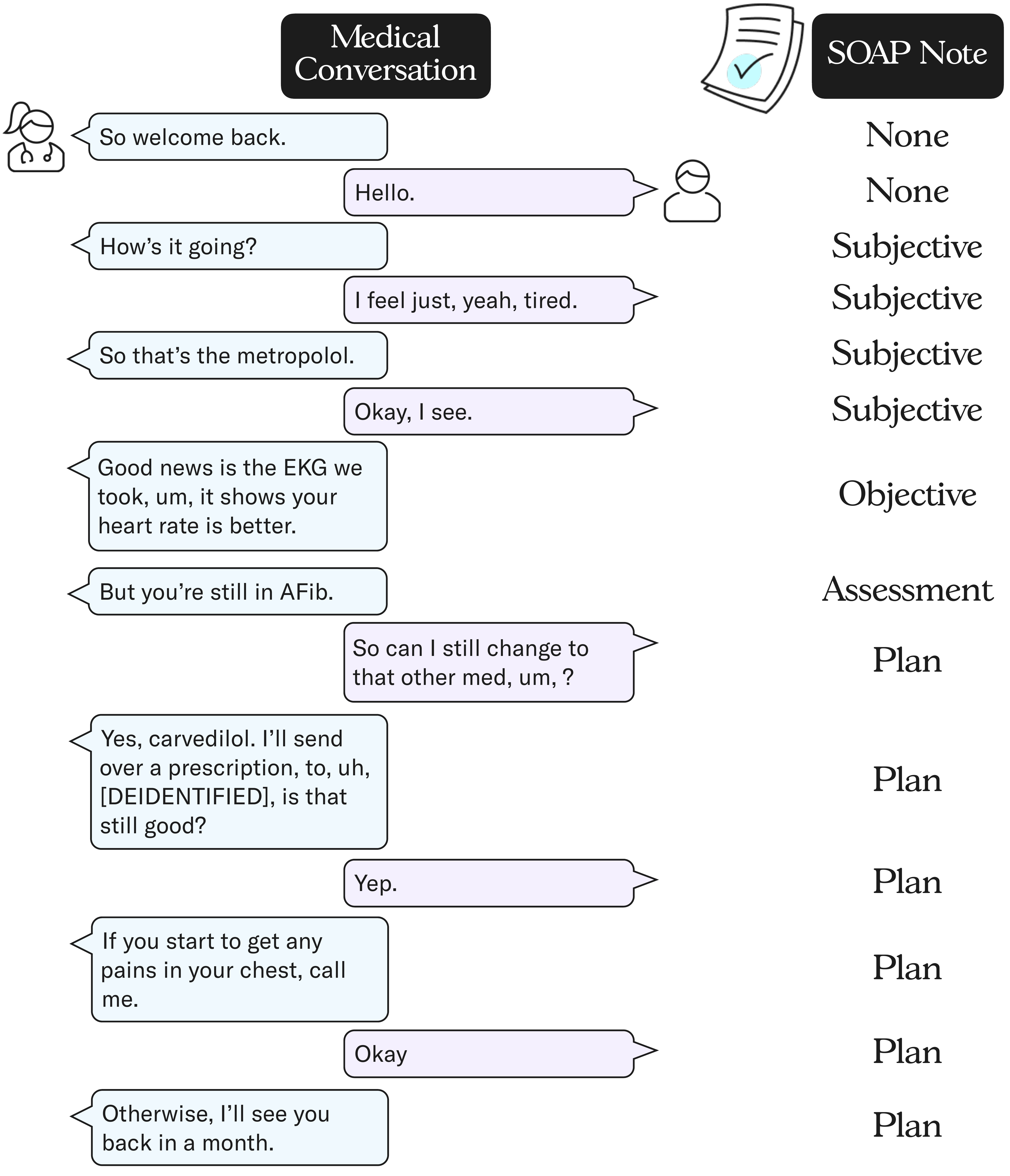}
\caption{A fictitious doctor-patient conversation. Each utterance is classified into one of the four SOAP note sections, or None.} 
\label{fig:note_example}
\end{figure*}

Figure \ref{fig:note_example} illustrates an example of a fictitious doctor-patient conversation and the corresponding classification of each utterance into one of the four SOAP sections, or None.

\begin{table*}[h!]
\begin{center}
\scalebox{1.00}{
\begin{tabular}{lllccc}
\toprule
Protected attribute & Disadvantaged group &Advantaged group &\% of Total &Size disadv/adv (\%/\%)\\
\midrule
\multirow[t]{3}{*}{Race/Ethnicity} & Black      &White  &13/77 &152,044/885,263\phantom{1,} (15\%/85\%)\\
& Hispanic &White &\phantom{0}6/77 &\phantom{0}66,729/885,263\phantom{1,} \phantom{0}(7\%/93\%) \\
    & Asian      &White &\phantom{0}3/77 &\phantom{0}36,497/885,263 \phantom{1,}\phantom{0}(4\%/96\%)\vspace{0.2cm}\\ 
\multirow[t]{2}{*}{Gender} &Female (Patient) & Male (Patient) &55/45 &635,376/512,368\phantom{1,} (55\%/45\%)\\
     &Female (Physician) & Male (Physician) &22/78 &248,366/900,223\phantom{1,} (22\%/78\%)\vspace{0.2cm}\\
\multirow[t]{2}{*}{Race/Ethn.+Gender}&Black female &White male &\phantom{0}8/35 &\phantom{0}87,779/398,528\phantom{1,} (18\%/82\%)\\
&Hispanic female &White male &\phantom{0}3/35 &\phantom{0}39,942/398,528\phantom{1,} \phantom{0}(9\%/91\%)\vspace{0.2cm}\\
\multirow[t]{6}{*}{Socio-economic status} 
&Unemployed &Full-time job &12/29 &134,666/328,500\phantom{1,} (29\%/71\%)\\
&Retired &Full-time job &35/29 &398,910/328,500\phantom{1,} (55\%/45\%)\\
&Nursing home & Living at home &\phantom{0}2/94 &\phantom{0}21,120/1,083,642 \phantom{0}(2\%/98\%)\\
&Incarcerated &Living at home &\phantom{0}1/94 &\phantom{00}7,970/1,083,642 \phantom{0}(1\%/99\%)\\
&Medicaid &Private insurance &\phantom{0}9/15 &103,260/176,920\phantom{1,} (37\%/63\%)\\
&Uninsured &Private insurance &\phantom{0}1/15 &\phantom{0}12,518/176,920\phantom{1,} \phantom{0}(7\%/93\%)\vspace{0.2cm}\\
\multirow[t]{2}{*}{Age} &Youth (0-17) &Adult (18-64)&\phantom{0}4/58 &\phantom{0}40,478/660,750 \phantom{1,}\phantom{0}(6\%/94\%)\\
&Older adult (65-98) &Adult (18-64) &38/58 &431,843/660,750\phantom{1,} (40\%/60\%)\vspace{0.2cm}\\
Obesity &$>=$ 250lbs. &$<$ 250lbs.&\phantom{0}8/84 &\phantom{0}97,576/962,080\phantom{1,}\phantom{0}(9\%/91\%)\vspace{0.2cm}\\
Mental health &Psychiatrist (Physician) &Other specialty (Physician) &\phantom{0}8/92 &\phantom{0}86,643/1,061,946\phantom{0}(8\%/92\%)\vspace{0.2cm}\\
Location &Other U.S. state &FL, CA, and NY &\phantom{0}68/32 &785,731/362,858\phantom{1,} (68\%/32\%)\\
\bottomrule
\end{tabular}}
\caption{{\small Protected attributes and their respective disadvantaged and advantaged groups. \% of Total is the percentage each group forms of the entire validation set; these may not sum to 100\% as not every conversation is tagged with every protected attribute. Size is the size of each group within this subset.}}
\label{tab:fair_groups}
\end{center}
\end{table*}

\section{Appendix B: Dataset}
\label{sec:appendix_b}
Table \ref{tab:fair_groups} summarizes all the protected attributes, disadvantaged and advantaged groups that were analyzed, with additional statistics describing their size relative proportions.

\section{Appendix C: Results}
\label{sec:appendix_c}
Table \ref{tab:all_results_1} and Table \ref{tab:all_results_2} list results across all groups, all SOAP note sections and all metrics.

\begin{table*}[!h] 
\begin{center} 
\scalebox{1.00}{ 
\begin{tabular}{p{2cm}p{2.4cm}p{2.4cm}p{1.3cm}p{1.5cm}p{1.5cm}p{1.5cm}} 
\toprule 
Protected \phantom{0} attribute & Disadvantaged group &Advantaged group &SOAP section &\texttt{AOD} [-0.1,0.1] &\texttt{EOR} [0.8,1.25] &\texttt{FORR} [0.8,1.25] \\
\midrule 
 \multirow[t]{15}{*}{Race/Ethn.} & \multirow[t]{5}{*}{Black} & \multirow[t]{5}{*}{White} &Subjective &\textcolor{ForestGreen}{
\phantom{-}0.02} &\textcolor{ForestGreen}{1.03} &\textcolor{ForestGreen}{0.96}\\
 & & &Objective &\textcolor{ForestGreen}{\phantom{-}0.01} &\textcolor{ForestGreen}{1.03} &\textcolor{ForestGreen}{1.10}\\
 & & &Assessment &\textcolor{ForestGreen}{-0.02} &\textcolor{ForestGreen}{0.94} &\textcolor{ForestGreen}{0.92}\\
 & & &Plan &\textcolor{ForestGreen}{\phantom{-}0.02} &\textcolor{ForestGreen}{1.02} &\textcolor{ForestGreen}{0.91}\\
 & & &None &\textcolor{ForestGreen}{-0.04} &\textcolor{red}{0.78} &\textcolor{ForestGreen}{1.01}\\
 & \multirow[t]{5}{*}{Hispanic} & \multirow[t]{5}{*}{White} &Subjective &\textcolor{ForestGreen}{\phantom{-}0.01} &\textcolor{
ForestGreen}{1.01} &\textcolor{ForestGreen}{0.98}\\
 & & &Objective &\textcolor{ForestGreen}{\phantom{-}0.02} &\textcolor{ForestGreen}{1.03} &\textcolor{ForestGreen}{1.16}\\
 & & &Assessment &\textcolor{ForestGreen}{-0.01} &\textcolor{ForestGreen}{0.98} &\textcolor{ForestGreen}{0.99}\\
 & & &Plan &\textcolor{ForestGreen}{\phantom{-}0.00} &\textcolor{ForestGreen}{0.99} &\textcolor{ForestGreen}{0.83}\\
 & & &None &\textcolor{ForestGreen}{-0.02} &\textcolor{ForestGreen}{0.88} &\textcolor{ForestGreen}{1.00}\\
 & \multirow[t]{5}{*}{Asian} & \multirow[t]{5}{*}{White} &Subjective &\textcolor{ForestGreen}{-0.03} &\textcolor{ForestGreen}{
0.93} &\textcolor{ForestGreen}{0.85}\\
 & & &Objective &\textcolor{ForestGreen}{-0.01} &\textcolor{ForestGreen}{0.99} &\textcolor{ForestGreen}{0.83}\\
 & & &Assessment &\textcolor{ForestGreen}{\phantom{-}0.00} &\textcolor{ForestGreen}{1.03} &\textcolor{ForestGreen}{1.06}\\
 & & &Plan &\textcolor{ForestGreen}{\phantom{-}0.00} &\textcolor{ForestGreen}{0.97} &\textcolor{red}{0.69}\\
 & & &None &\textcolor{ForestGreen}{-0.01} &\textcolor{ForestGreen}{0.94} &\textcolor{ForestGreen}{1.03}\\
 \multirow[t]{15}{*}{Gender} & \multirow[t]{5}{*}{Female (Patient)} & \multirow[t]{5}{*}{Male (Patient)} &Subjective &\textcolor{ForestGreen}{\phantom{-}0.00} &\textcolor{ForestGreen}{1.00} &\textcolor{ForestGreen}{0.99}\\
 & & &Objective &\textcolor{ForestGreen}{-0.01} &\textcolor{ForestGreen}{0.99} &\textcolor{ForestGreen}{1.16}\\
 & & &Assessment &\textcolor{ForestGreen}{\phantom{-}0.01} &\textcolor{ForestGreen}{1.02} &\textcolor{ForestGreen}{0.94}\\
 & & &Plan &\textcolor{ForestGreen}{-0.01} &\textcolor{ForestGreen}{0.98} &\textcolor{ForestGreen}{0.98}\\
 & & &None &\textcolor{ForestGreen}{\phantom{-}0.00} &\textcolor{ForestGreen}{0.98} &\textcolor{ForestGreen}{1.00}\\
 & \multirow[t]{5}{*}{Female (Physician)} & \multirow[t]{5}{*}{Male (Physician)} &Subjective &\textcolor{ForestGreen}{-0.02} &
\textcolor{ForestGreen}{0.95} &\textcolor{ForestGreen}{1.02}\\
 & & &Objective &\textcolor{ForestGreen}{-0.01} &\textcolor{ForestGreen}{0.97} &\textcolor{ForestGreen}{0.97}\\
 & & &Assessment &\textcolor{ForestGreen}{\phantom{-}0.01} &\textcolor{ForestGreen}{1.05} &\textcolor{ForestGreen}{1.14}\\
 & & &Plan &\textcolor{ForestGreen}{-0.01} &\textcolor{ForestGreen}{0.98} &\textcolor{ForestGreen}{1.02}\\
 & & &None &\textcolor{ForestGreen}{\phantom{-}0.02} &\textcolor{ForestGreen}{1.15} &\textcolor{ForestGreen}{0.97}\\
 \multirow[t]{15}{*}{Race+Gender} & \multirow[t]{5}{*}{Black female} & \multirow[t]{5}{*}{White male} &Subjective &\textcolor{ForestGreen}{\phantom{-}0.02} &\textcolor{ForestGreen}{1.04} &\textcolor{ForestGreen}{0.98}\\
 & & &Objective &\textcolor{ForestGreen}{\phantom{-}0.01} &\textcolor{ForestGreen}{1.02} &\textcolor{ForestGreen}{1.17}\\
 & & &Assessment &\textcolor{ForestGreen}{-0.01} &\textcolor{ForestGreen}{0.94} &\textcolor{ForestGreen}{0.88}\\
 & & &Plan &\textcolor{ForestGreen}{\phantom{-}0.02} &\textcolor{ForestGreen}{1.01} &\textcolor{ForestGreen}{0.93}\\
 & & &None &\textcolor{ForestGreen}{-0.04} &\textcolor{red}{0.76} &\textcolor{ForestGreen}{1.01}\\
 & \multirow[t]{5}{*}{Hispanic female} & \multirow[t]{5}{*}{White male} &Subjective &\textcolor{ForestGreen}{\phantom{-}0.01} 
&\textcolor{ForestGreen}{1.00} &\textcolor{ForestGreen}{0.98}\\
 & & &Objective &\textcolor{ForestGreen}{\phantom{-}0.02} &\textcolor{ForestGreen}{1.06} &\textcolor{blue}{1.66}\\
 & & &Assessment &\textcolor{ForestGreen}{\phantom{-}0.02} &\textcolor{ForestGreen}{1.03} &\textcolor{ForestGreen}{0.89}\\
 & & &Plan &\textcolor{ForestGreen}{-0.01} &\textcolor{ForestGreen}{0.95} &\textcolor{red}{0.71}\\
 & & &None &\textcolor{ForestGreen}{-0.03} &\textcolor{ForestGreen}{0.82} &\textcolor{ForestGreen}{1.02}\\
 \multirow[t]{20}{*}{Socio-economic} & \multirow[t]{5}{*}{Unemployed} & \multirow[t]{5}{*}{Full-time} &Subjective &\textcolor{ForestGreen}{\phantom{-}0.03} &\textcolor{ForestGreen}{1.03} &\textcolor{ForestGreen}{0.88}\\
 & & &Objective &\textcolor{ForestGreen}{-0.02} &\textcolor{ForestGreen}{0.97} &\textcolor{ForestGreen}{1.16}\\
 & & &Assessment &\textcolor{ForestGreen}{-0.01} &\textcolor{ForestGreen}{0.97} &\textcolor{ForestGreen}{0.96}\\
 & & &Plan &\textcolor{ForestGreen}{-0.01} &\textcolor{ForestGreen}{1.00} &\textcolor{ForestGreen}{1.19}\\
 & & &None &\textcolor{ForestGreen}{\phantom{-}0.00} &\textcolor{ForestGreen}{0.97} &\textcolor{ForestGreen}{1.02}\\
 & \multirow[t]{5}{*}{Retired} & \multirow[t]{5}{*}{Full-time} &Subjective &\textcolor{ForestGreen}{-0.01} &\textcolor{ForestGreen}{0.98} &\textcolor{ForestGreen}{1.12}\\
 & & &Objective &\textcolor{ForestGreen}{\phantom{-}0.00} &\textcolor{ForestGreen}{1.01} &\textcolor{ForestGreen}{1.22}\\
 & & &Assessment &\textcolor{ForestGreen}{\phantom{-}0.06} &\textcolor{ForestGreen}{1.13} &\textcolor{ForestGreen}{0.95}\\
 & & &Plan &\textcolor{ForestGreen}{\phantom{-}0.00} &\textcolor{ForestGreen}{1.01} &\textcolor{ForestGreen}{1.05}\\
 & & &None &\textcolor{ForestGreen}{-0.01} &\textcolor{ForestGreen}{0.92} &\textcolor{ForestGreen}{0.97}\\
 & \multirow[t]{5}{*}{Nursing home} & \multirow[t]{5}{*}{Living at home} &Subjective &\textcolor{ForestGreen}{\phantom{-}0.05}
 &\textcolor{ForestGreen}{1.05} &\textcolor{ForestGreen}{0.9}\\
 & & &Objective &\textcolor{ForestGreen}{-0.01} &\textcolor{ForestGreen}{1.00} &\textcolor{blue}{2.35}\\
 & & &Assessment &\textcolor{ForestGreen}{-0.03} &\textcolor{ForestGreen}{0.88} &\textcolor{ForestGreen}{0.96}\\
 & & &Plan &\textcolor{ForestGreen}{-0.02} &\textcolor{ForestGreen}{1.01} &\textcolor{blue}{1.32}\\
 & & &None &\textcolor{ForestGreen}{\phantom{-}0.02} &\textcolor{ForestGreen}{1.03} &\textcolor{ForestGreen}{0.98}\\
 & \multirow[t]{5}{*}{Incarcerated} & \multirow[t]{5}{*}{Living at home} &Subjective &\textcolor{ForestGreen}{-0.04} &\textcolor{ForestGreen}{0.93} &\textcolor{ForestGreen}{1.02}\\
 & & &Objective &\textcolor{red}{-0.17}* &\textcolor{red}{0.63}* &\textcolor{red}{0.75}*\\
 & & &Assessment &\textcolor{blue}{\phantom{-}0.18} &\textcolor{blue}{1.38} &\textcolor{ForestGreen}{0.94}\\
 & & &Plan &\textcolor{ForestGreen}{-0.05} &\textcolor{ForestGreen}{0.95} &\textcolor{red}{0.71}\\
 & & &None &\textcolor{ForestGreen}{-0.04} &\textcolor{ForestGreen}{0.81} &\textcolor{ForestGreen}{1.00}\\
\bottomrule  
\end{tabular}}  
\caption{{\small Group disparities metrics, organized by protected attribute and SOAP note section. Numbers in green are within the thresholds, red are disparate favoring the advantaged group, blue favoring the disadvantaged group. Numbers with an asterisk indicate the size of the group was too small to analyze (75 utterances). \texttt{AOD}=average odds difference, \texttt{EOR}=equal opportunity ratio, \texttt{FORR}=false omission rate ratio.}}
\label{tab:all_results_1} 
\end{center}
\end{table*}

\begin{table*}[!h] 
\begin{center} 
\scalebox{1.00}{ 
\begin{tabular}{p{2cm}p{2.4cm}p{2.4cm}p{1.3cm}p{1.5cm}p{1.5cm}p{1.5cm}} 
\toprule 
Protected \phantom{0} attribute & Disadvantaged group &Advantaged group &SOAP section &\texttt{AOD} [-0.1,0.1] &\texttt{EOR} [0.8,1.25] &\texttt{FORR} [0.8,1.25] \\
\midrule 
\multirow[t]{10}{*}{Socio-economic} & \multirow[t]{5}{*}{Medicaid} & \multirow[t]{5}{*}{Private insurance} &Subjective &\textcolor{ForestGreen}{\phantom{-}0.04} 
&\textcolor{ForestGreen}{1.08} &\textcolor{ForestGreen}{1.02}\\
 & & &Objective &\textcolor{ForestGreen}{-0.01} &\textcolor{ForestGreen}{0.98} &\textcolor{ForestGreen}{1.23}\\
 & & &Assessment &\textcolor{ForestGreen}{-0.01} &\textcolor{ForestGreen}{0.97} &\textcolor{ForestGreen}{0.96}\\
 & & &Plan &\textcolor{ForestGreen}{\phantom{-}0.01} &\textcolor{ForestGreen}{1.05} &\textcolor{blue}{1.69}\\
 & & &None &\textcolor{ForestGreen}{-0.01} &\textcolor{ForestGreen}{0.97} &\textcolor{ForestGreen}{1.00}\\
 & \multirow[t]{5}{*}{Uninsured} & \multirow[t]{5}{*}{Private insurance} &Subjective &\textcolor{ForestGreen}{\phantom{-}0.04}
 &\textcolor{ForestGreen}{1.07} &\textcolor{ForestGreen}{0.86}\\
 & & &Objective &\textcolor{ForestGreen}{-0.03} &\textcolor{ForestGreen}{0.94} &\textcolor{red}{0.67}\\
 & & &Assessment &\textcolor{ForestGreen}{-0.01} &\textcolor{ForestGreen}{0.99} &\textcolor{ForestGreen}{1.11}\\
 & & &Plan &\textcolor{ForestGreen}{-0.01} &\textcolor{ForestGreen}{0.99} &\textcolor{ForestGreen}{0.82}\\
 & & &None &\textcolor{ForestGreen}{-0.02} &\textcolor{ForestGreen}{0.96} &\textcolor{ForestGreen}{1.07}\\
 \multirow[t]{15}{*}{Age} & \multirow[t]{5}{*}{Youth} & \multirow[t]{5}{*}{Adult} &Subjective &\textcolor{ForestGreen}{-0.02} 
&\textcolor{ForestGreen}{0.91} &\textcolor{ForestGreen}{1.01}\\
 & & &Objective &\textcolor{ForestGreen}{-0.06} &\textcolor{ForestGreen}{0.87} &\textcolor{blue}{1.43}\\
 & & &Assessment &\textcolor{ForestGreen}{\phantom{-}0.00} &\textcolor{ForestGreen}{0.99} &\textcolor{ForestGreen}{0.97}\\
 & & &Plan &\textcolor{ForestGreen}{-0.04} &\textcolor{ForestGreen}{0.91} &\textcolor{ForestGreen}{1.01}\\
 & & &None &\textcolor{ForestGreen}{\phantom{-}0.04} &\textcolor{ForestGreen}{1.15} &\textcolor{ForestGreen}{0.93}\\
 & \multirow[t]{5}{*}{Older adult} & \multirow[t]{5}{*}{Adult} &Subjective &\textcolor{ForestGreen}{\phantom{-}0.00} &\textcolor{ForestGreen}{0.98} &\textcolor{ForestGreen}{0.88}\\
 & & &Objective &\textcolor{ForestGreen}{\phantom{-}0.00} &\textcolor{ForestGreen}{0.99} &\textcolor{red}{0.69}\\
 & & &Assessment &\textcolor{ForestGreen}{-0.06} &\textcolor{ForestGreen}{0.89} &\textcolor{ForestGreen}{1.09}\\
 & & &Plan &\textcolor{ForestGreen}{\phantom{-}0.00} &\textcolor{ForestGreen}{0.98} &\textcolor{ForestGreen}{0.81}\\
 & & &None &\textcolor{ForestGreen}{\phantom{-}0.02} &\textcolor{ForestGreen}{1.11} &\textcolor{ForestGreen}{1.03}\\
 \multirow[t]{15}{*}{Obesity} & \multirow[t]{5}{*}{$>=$250lbs.} & \multirow[t]{5}{*}{$<$250lbs.} &Subjective &\textcolor{ForestGreen}{\phantom{-}0.01} &\textcolor{ForestGreen}{1.02} &\textcolor{ForestGreen}{1.01}\\
 & & &Objective &\textcolor{ForestGreen}{\phantom{-}0.01} &\textcolor{ForestGreen}{1.02} &\textcolor{ForestGreen}{1.09}\\
 & & &Assessment &\textcolor{ForestGreen}{-0.01} &\textcolor{ForestGreen}{0.96} &\textcolor{ForestGreen}{0.88}\\
 & & &Plan &\textcolor{ForestGreen}{\phantom{-}0.02} &\textcolor{ForestGreen}{1.04} &\textcolor{ForestGreen}{1.23}\\
 & & &None &\textcolor{ForestGreen}{-0.01} &\textcolor{ForestGreen}{0.96} &\textcolor{ForestGreen}{1.03}\\
 \multirow[t]{15}{*}{Mental health} & \multirow[t]{5}{*}{Psychiatrist} & \multirow[t]{5}{*}{Other specialty} &Subjective &\textcolor{ForestGreen}{-0.02} &\textcolor{ForestGreen}{0.93} &\textcolor{ForestGreen}{0.97}\\
 & & &Objective &\textcolor{ForestGreen}{-0.05} &\textcolor{ForestGreen}{0.93} &\textcolor{blue}{4.15}\\
 & & &Assessment &\textcolor{ForestGreen}{\phantom{-}0.08} &\textcolor{ForestGreen}{1.16} &\textcolor{ForestGreen}{1.02}\\
 & & &Plan &\textcolor{ForestGreen}{-0.04} &\textcolor{ForestGreen}{1.00} &\textcolor{blue}{1.54}\\
 & & &None &\textcolor{ForestGreen}{\phantom{-}0.06} &\textcolor{blue}{1.30} &\textcolor{ForestGreen}{0.97}\\
 \multirow[t]{15}{*}{Location} & \multirow[t]{5}{*}{Other U.S. state} & \multirow[t]{5}{*}{FL, CA, and NY} &Subjective &\textcolor{ForestGreen}{-0.01} &\textcolor{ForestGreen}{0.98} &\textcolor{ForestGreen}{0.96}\\
 & & &Objective &\textcolor{ForestGreen}{\phantom{-}0.00} &\textcolor{ForestGreen}{1.00} &\textcolor{ForestGreen}{1.17}\\
 & & &Assessment &\textcolor{ForestGreen}{\phantom{-}0.01} &\textcolor{ForestGreen}{1.03} &\textcolor{ForestGreen}{1.08}\\
 & & &Plan &\textcolor{ForestGreen}{-0.01} &\textcolor{ForestGreen}{0.99} &\textcolor{ForestGreen}{0.98}\\
 & & &None &\textcolor{ForestGreen}{\phantom{-}0.01} &\textcolor{ForestGreen}{1.10} &\textcolor{ForestGreen}{0.99}\\
 \bottomrule  
\end{tabular}}  
\caption{{\small (Continued) Group disparities metrics, organized by protected attribute and SOAP note section. Numbers in green are within the thresholds, red are disparate favoring the advantaged group, blue favoring the disadvantaged group. \texttt{AOD}=average odds difference, \texttt{EOR}=equal opportunity ratio, \texttt{FORR}=false omission rate ratio.}}
\label{tab:all_results_2} 
\end{center}
\end{table*}

\end{document}